\documentclass[12pt,preprint]{aastex63}
\usepackage{graphicx}

\received{...}
\revised{...}
\accepted{...}
\submitjournal{ApJ}

\begin{document}

\title{The Details of Limb Brightening Reveal the Structure of the Base of the Jet in M\,87 for the First Time}

\correspondingauthor{Brian Punsly}
\email{brian.punsly@cox.net}

\author{Brian Punsly}
\affiliation{1415 Granvia Altamira, Palos Verdes Estates CA, USA 90274}
\affiliation{ICRANet, Piazza della Repubblica 10 Pescara 65100, Italy}
\affiliation{ICRA, Physics Department, University La Sapienza, Roma, Italy}

\begin{abstract}
It has become commonplace is astronomy to describe the transverse coarse structure of jets in loosely defined terms such as ``sheath" and ``spine" based on discussions of parsec scale properties. But, the applicability, dimension and prominence of these features on sub-lt-yr scales has previously been unconstrained by observation. The first direct evidence of jet structure near the source in M\,87 is extreme limb brightening (a double-rail morphology), 0.3 - 0.6 mas from the source, that is prominent in observations with high resolution and sensitivity.  Intensity cross-cuts of these images provide three strong, interdependent constraints on the geometry responsible for the double-rail morphology: the rail to rail separation, the peak to trough intensity ratio and the rail widths. Analyzing these constraints indicates that half or more of the jet volume resides in a thick-walled, tubular, mildly relativistic, protonic jet only $\sim 0.25$ lt-yr (or $\sim 300$ M, where M is the central black hole mass in geometrized units) from the source. By contrast, the Event Horizon Telescope Collaboration interprets their observations with the aid of general relativistic magnetohydrodynamic simulations that produce an invisible (by construction) jet with a surrounding luminous, thin sheath. Yet, it is shown that synthetic images of simulated jets are center brightened 0.3 - 0.6 mas from the source. This serious disconnection with observation occurs in a region previously claimed in the literature to be well represented by the simulations. The limb brightening analysis motivates a discussion of possible simulation modifications to improve conformance with observations.
\end{abstract}


\section{Introduction}

  The powerful jet in the nearby galaxy M\,87 ($\approx 16.8$ Mpc distant) is the best candidate for observing jet launching and the jet base in detail, therefore a prime target of $\sim$1.3mm VLBI (Very Long Baseline Interferometry), the Event Horizon Telescope (EHT)\citep{doe12}. The EHT has not yet detected a jet, only an annulus of emission \citep{eht19}. The observations are very challenging and are interpreted with the aid of a library of general relativistic magnetohydrodynamic (GRMHD) numerical simulations \citep{por19}. There is little observational information on the jet configuration just above the EHT annulus and therefore little to restrict the jet produced in the GRMHD simulations. To improve this situation, this study explores the extreme limb brightening in the highest sensitivity 3.5mm and 7mm VLBI experiments in order to quantify the dimensions and composition of the transverse structure $\sim 0.25-0.45$ lt-yrs downstream of the EHT annulus.

\par The highest signal to noise and sensitivity 43(86) GHz VLBI image is reproduced in the top (bottom) left hand panel of Figure 1. The details of the 43(86) GHz image and the observations have been discussed previously in \citet{wal18,pun21} and \citep{had16,kim18}. The images are formatted in a manner that accentuates the jet's transverse structure. The contour levels are chosen to highlight the jet at the expense of not capturing a much brighter core and an unresolved region of the jet. The images are rotated so the jet axis, z, is horizontal, i.e. a rotation of PA=$-23^{\circ}$ \citep{had16}.  The elliptical Gaussian restoring beam (the blue ellipses) have full width at half maximum, FWHM, 0.16 mas x 0.21 mas, PA=$0^{\circ}$ and 0.116 mas x 0.307 mas, PA=$-9^{\circ}$ at 43 and 86 GHz, respectively. These are also rotated by PA=$-23^{\circ}$ in Figure 1. The 86 GHz image in Figure 1 is the \citet{kim18} reconstruction of the radio source that was observed and reconstructed originally in \citep{had16}.  Assuming intrinsic bilateral symmetry, it was deduced from the 43 GHz image that the associated Doppler boosting (with a line of sight (LOS) of $18^{\circ}$ to the jet axis \citet{eht19,cha19}) corresponds to mildly relativistic speeds for the bulk velocity of the jet, accelerating from $\sim 0.27$c (z= 0.4 mas) to $\sim 0.38$c (z=0.65 mas)\citep{pun21}. The bright ridges (``limb brightening" in the sky plane) are the main topic of this study. An intensity cross-cut at z = 0.5 mas is reproduced in the right hand panel of Figure 1 \citep{pun22}. Notice that the peak to trough intensity ratios are very large $\sim 6-10$ at 43 GHz. The figure also compares the width of the ridges (black double Gaussian fit) to the Gaussian restoring beam projected along the cross-cut (dashed green). There are 6 cross-cuts for a total of 12 ridges in \citep{pun22}. In that paper the restoring beam was de-convolved from the black Gaussian fits. The 12 de-convolved ridge FWHM were on average 42\% of the jet radius, $R$. Ostensibly, this indicates intrinsically wide ridges. Wide intensity ridges spatially juxtaposed to a central deep trough are two behaviors that seem to be at odds with each other. The intrinsic configuration is obfuscated by the finite resolution of VLBI and a nearly polar LOS. The main goal of this study is to understand the intrinsic source geometry that is compatible with the detailed structure of the intensity cross-cuts.
\begin{figure}
\begin{center}
\vspace{-3cm}
\includegraphics[width= 0.45\textwidth]{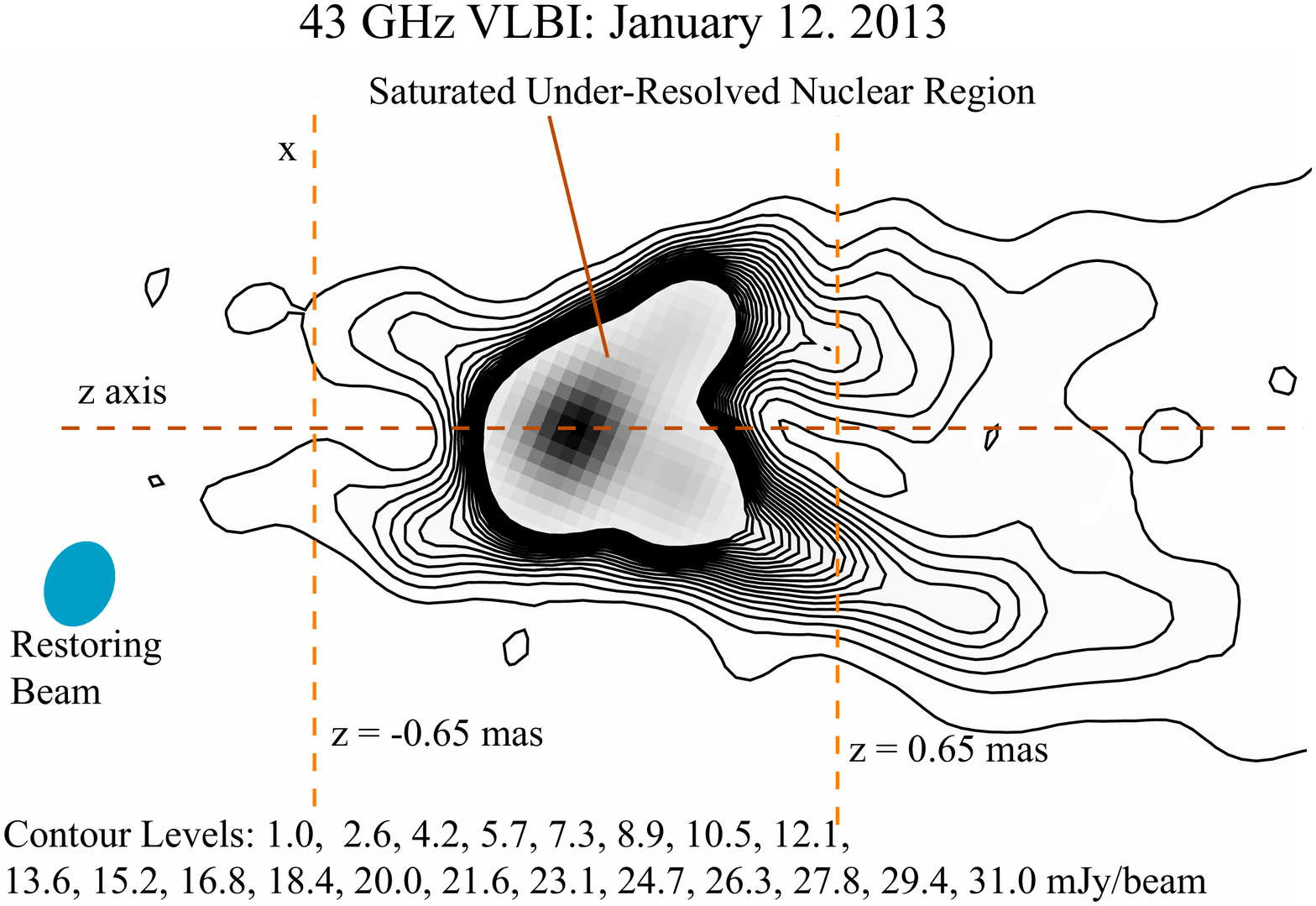}
\includegraphics[width= 0.45\textwidth]{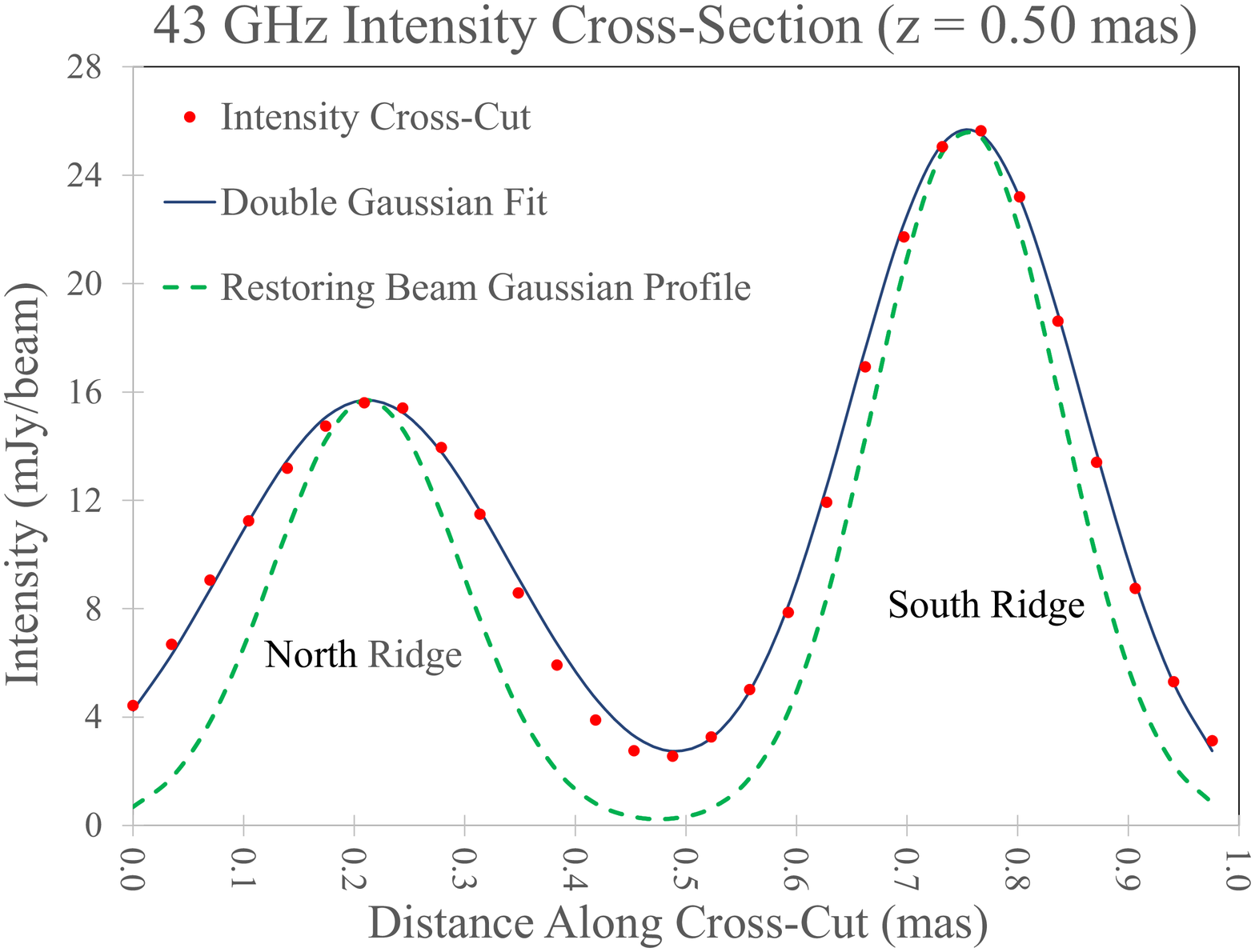}
\includegraphics[width= 0.44\textwidth]{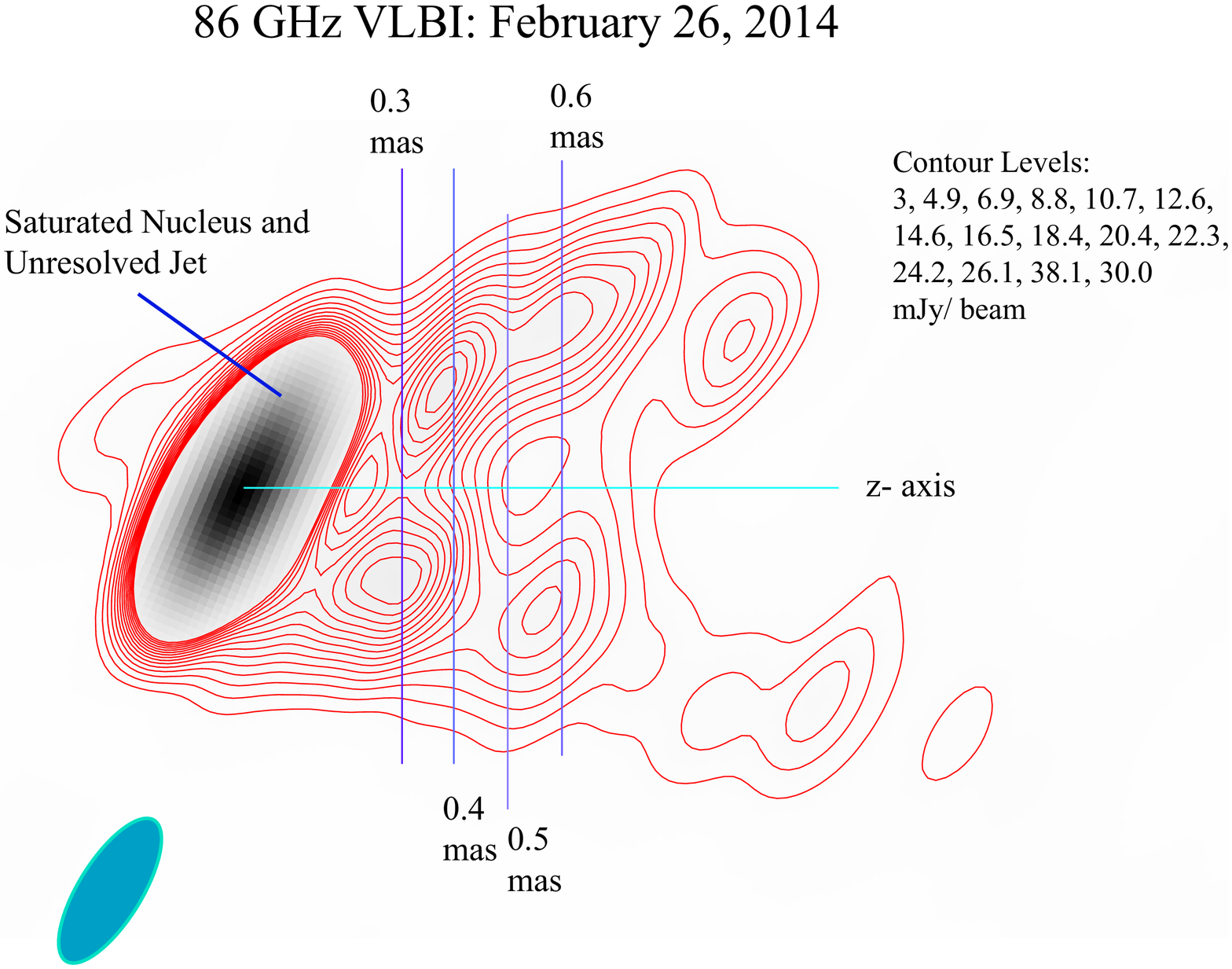}
\includegraphics[width= 0.45\textwidth]{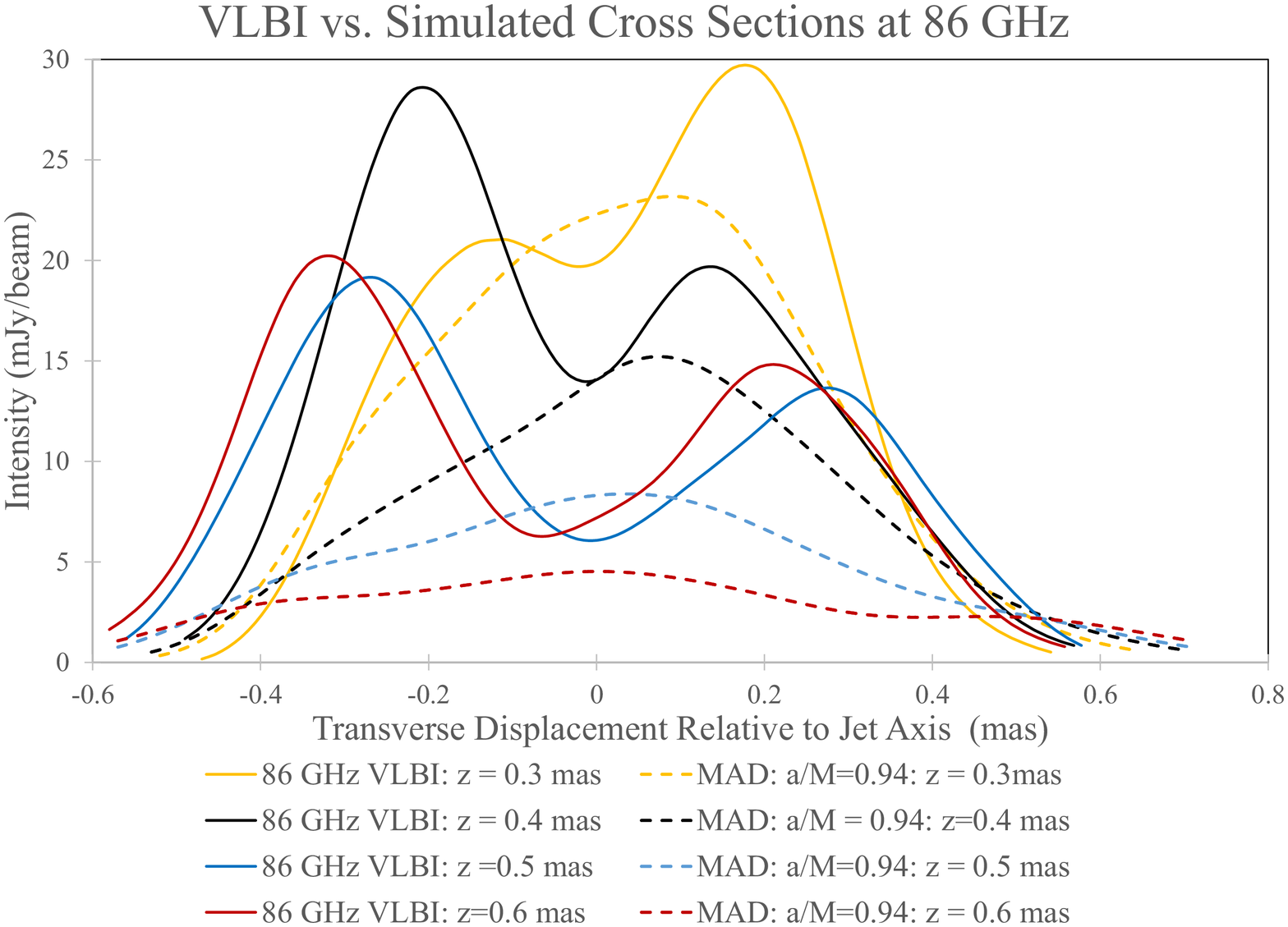}
\caption{\footnotesize{A high sensitivity 43 (86) GHz VLBI jet image is shown on the top (bottom) left. The limbs of the approaching jet in the 43 GHz image are very bright relative to a faint axial trough. The large degree of limb brightening is shown in the cross-cut on the top right. The bottom right hand frame compares intensity cross-cuts of a synthetic image in Figure 4 (dashed curves) from the most promising candidate simulation (``MAD: a/M=0.94") to cross-cuts of the VLBI image at 86 GHz (solid curves).}}
\end{center}
\end{figure}

\section{The Extreme Limb Brightening and Wide Jet Ridges}
Limb brightening is a phenomenon associated with astrophysical objects such as enhanced ovals and ridges of emission in planetary nebulae and supernovae remnants \citep{mas90,cor03,jon98}. These objects are relatively nearby and their dimension are easily resolved. With very high resolution, the peak to trough ratio diverges as $1/\sqrt{W}$ for a narrow ridge width, $W$ \citep{mas90}. But the limited resolution of VLBI is relatively insensitive to jet wall variations if $W$ is less than half of the beam FWHM (see top right of Figure 2). The other obfuscating circumstance is the nearly polar LOS and gradients in the emissivity along the jet. The LOS crosses the near side of the jet at low emissivity and exits the far side at high emissivity. In a cross-cut, the intensity trough includes contributions from LOS exit points from the jet wall that are closer to the nucleus than any other LOS and this is the region of highest emissivity (x=0 in bottom panel of Figure 2). This makes the very extreme limb brightening even more difficult to explain.

\begin{figure}
\begin{center}
\includegraphics[width= 0.47\textwidth]{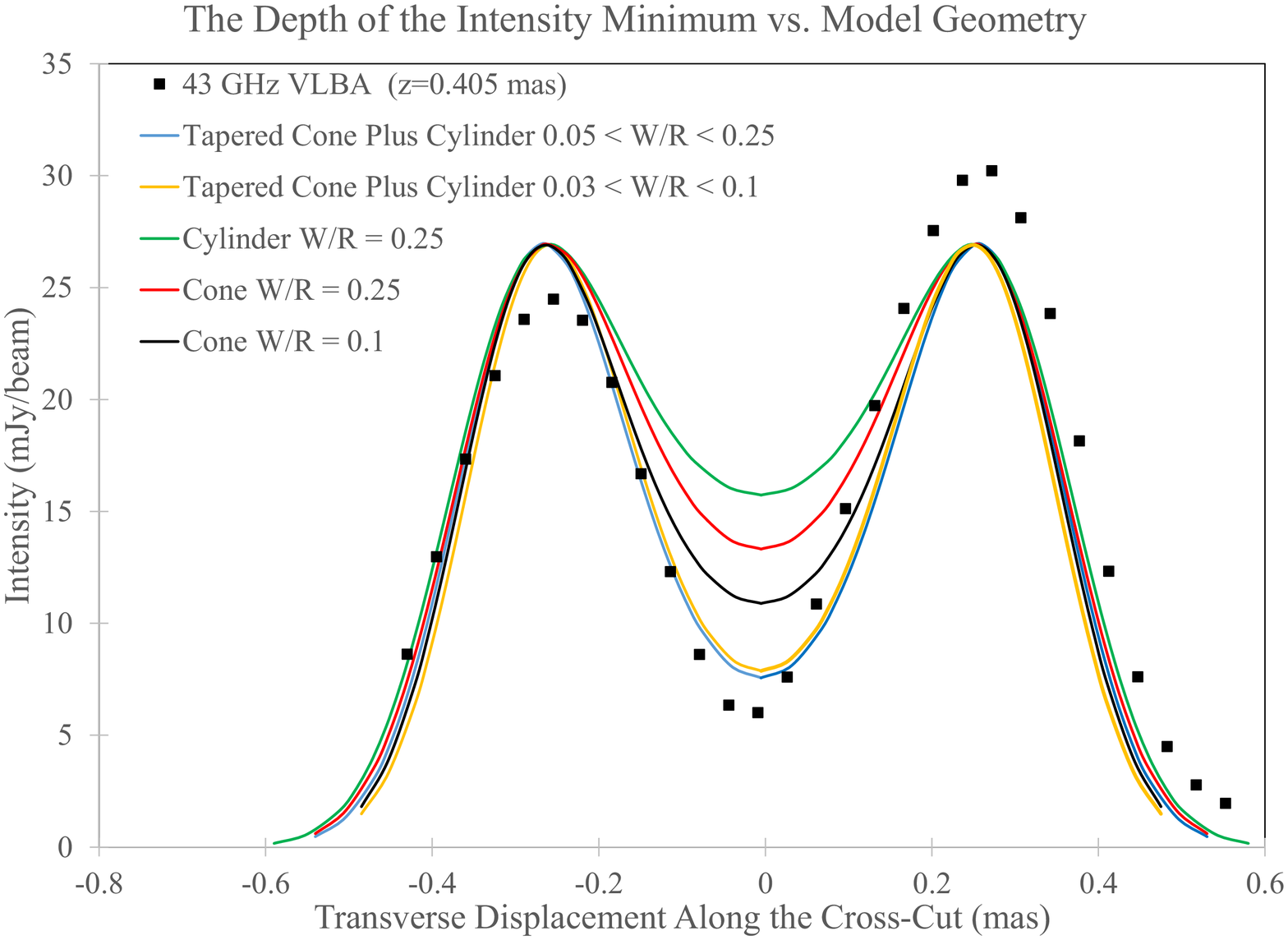}
\includegraphics[width= 0.47\textwidth]{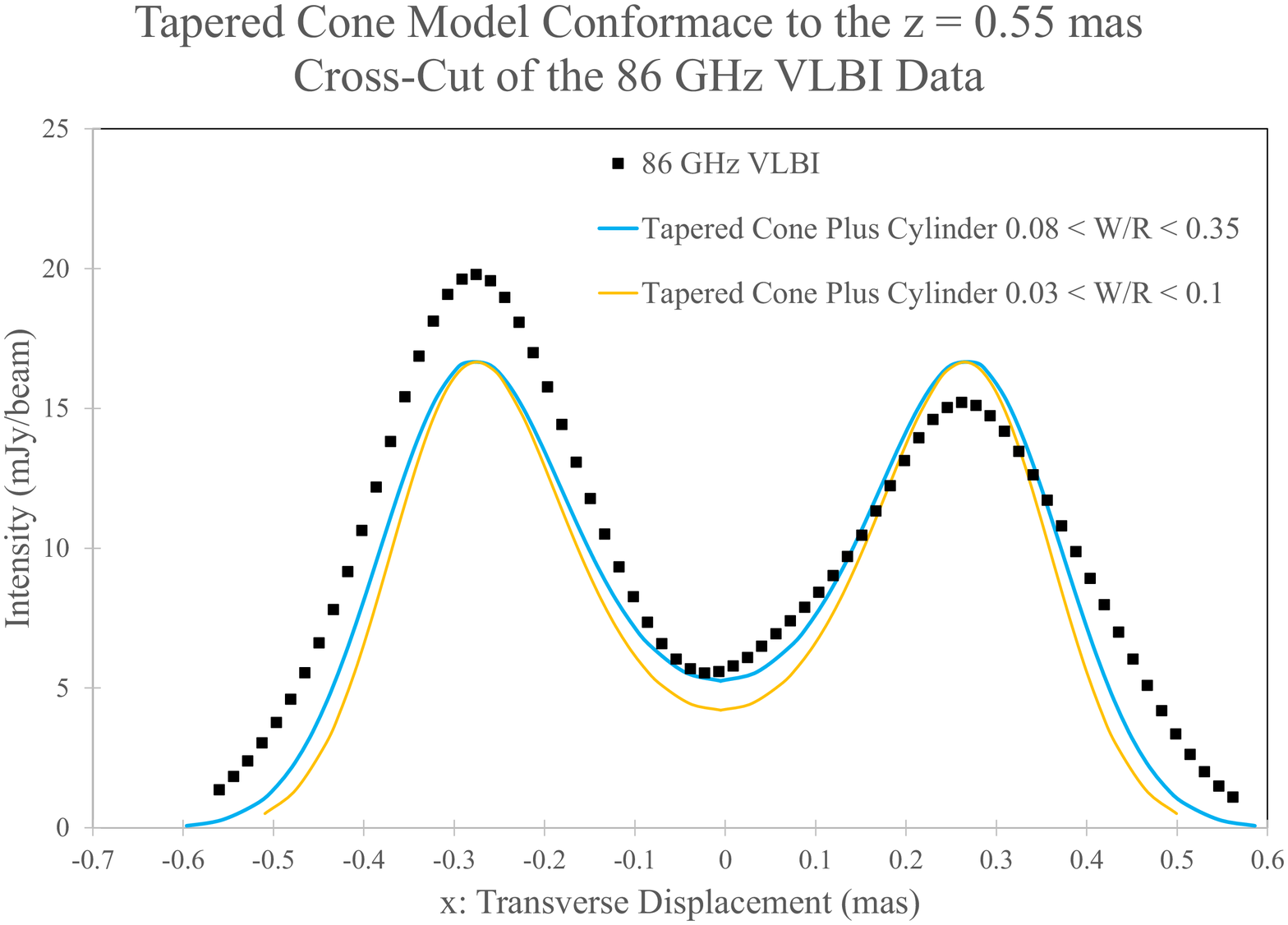}
\includegraphics[width= 0.6\textwidth]{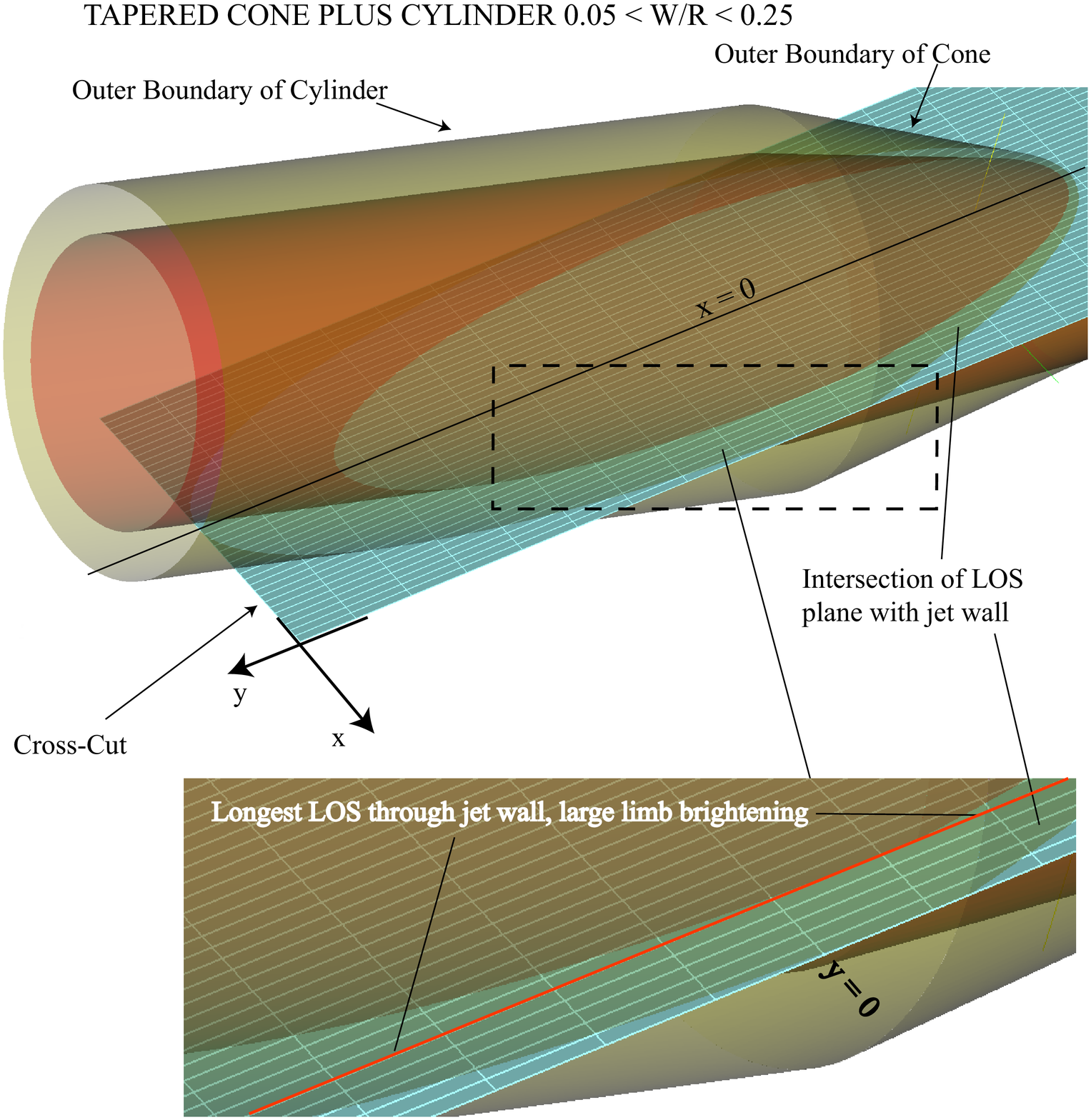}
\caption{\footnotesize{The top left panel is an experiment to explore ratio of the intensity peak to trough as a function of the geometrical model of the jet. The top right experiment explores the effect of wall thickness on the cross-cut. The nomenclature, $0.05 < W/R < 0.25$, is an abbreviation for a cone plus cylinder with $W/R=0.25$ at the junction of the cylinder and cone and tapers to $W/R=0.05$ when the LOS exits the cone wall at x=0. In order to clarify this, an example of a cone with tapered walls plus cylinder is shown in detail in the bottom panel. The image is a snapshot from a 3-D model in Paraview 3.3.0. The blue, (x, y), plane is foliated by the set of all LOS that constitute a cross-cut. The intersection of the plane with the tubular jet is in dark olive green. The key feature for extreme limb brightening is that the intersection is narrow at the exit of the LOS plane from the tubular wall on the right hand side. The dashed black rectangle shows a cutout of the region of maximum limb brightening that is magnified at the bottom. This region has the longest LOS through the tubular jet wall (red line) as indicated by the white grid overlayed on the blue LOS plane.}}
\end{center}
\end{figure}
\par The intensity cross-cuts provide three strong, interdependent constraints, the peak to peak separation, the peak to trough ratio and the ridge widths. Limb brightening was investigated with simplified axisymmetric geometrical models of emissivity that are not intended to replicate the actual jet. They are investigative tools to understand which geometrical properties can contribute to both the deep trough and the wide ridges with a minimal number of degrees of freedom. For $z< 1$ mas, it was previously found that $\alpha \approx 0.8$ from 43 GHz to 86 GHz, where the flux density, $S(\nu)\sim \nu^{-\alpha}$ \cite{had16}. Thus, the jet appears to be optically thin in the frequency range of interest. Therefore, absorption is ignored in the equation of radiative transfer. Each axisymmetric model is used to generate cross-cuts by numerically approximating the following convolution integral,
\begin{equation}
I(x_{o},\,z_{o})=\int{\int{\int{N\epsilon(x,y,z)}\frac{1}{2\pi \sqrt{\mid\mathbf{a}\mid \mid\mathbf{b}\mid}}\rm{e}^{{-\frac{\mid\mathbf{r}\cdot\mathbf{a}\mid^{2}}{2(\mathbf{a}\cdot \mathbf{a})^{2}}}-\frac{\mid\mathbf{r}\cdot\mathbf{b}\mid^{2}}{2(\mathbf{b}\cdot \mathbf{b})^{2}}}}}dy\,dx\,dz\;,\quad \epsilon(x,y,z) \approx \epsilon(y)=-0.781y +1\;.
\end{equation}
where, the restoring beam is defined by an elliptical Gaussian with major (minor) axis $\mathbf{a}$ ($\mathbf{b}$) of length of the standard deviation $\sigma$= FWHM/2.35, the displacement from $(x_{o}, z_{o})$ in the sky (or x-z) plane is $\mathbf{r}$ and the emissivity is $\epsilon$. $N$ is the only free parameter once a geometrical model is chosen and sets the normalization to agree with the peak intensity (note 0.1 mas $\approx 2.5 \times 10^{16}$ cm). The intensity, $I(x_{o},\,z_{o})$, is evaluated in a cross-section at $z=z_{o}$ at all points $x=x_{o}$ along the transverse slice. $y$ is orthogonal to the sky plane with LOS = $18^{\circ}$. Since $dy/dz' = \cos{18^{\circ}}\approx 1$ ($z'$ is measured along the symmetry axis of the jet), from Table 1, assuming a cross-cut at $z\sim 0.4$mas, we can estimate the emissivity in the axisymmetric approximation. The emissivity, $\epsilon(x,y,z) \approx\epsilon(y)$ is given in Equation (1), where the LOS intersects the jet wall, and is 0 elsewhere. The function is referenced to the point where the outer boundary of the slice through the jet wall by the LOS plane is tangent to the y axis as indicated in the bottom panel of Figure 2. At this point, $\epsilon(y)=1$ and $y=0$. This is emissivity in the observer's frame of reference, so it includes effects like Doppler boosting. $\epsilon(x,y,z)$ in Equation (1) is suitable for the purposes of this study because in the model analysis, the peak to trough ratios and the ridge width are relatively insensitive to the emissivity profile (compared to the geometry) unless the gradients increase by at least an order of magnitude from those indicated by Table 1.

\par Figure 2 highlights the results of an investigation of the geometrical properties that can produce deep surface brightness troughs between adjacent, wide intensity peaks. The top left panel investigates the effect of geometry on the depth of the trough. The top right panel explores the relationship between widening the jet walls, the intensity ridge width and the trough depth. The 43 GHz VLBI cross-cut in the top left panel was chosen for the investigation of the trough depth for two reasons. First, it is the most symmetric cross-cut from \citet{pun22} and the models are axisymmetric. Secondly, it is the only cross-cut from \citet{pun22} in which models were found that achieve the trough depth relative to the peak. Note that the trough in the top right hand panel of the Figure 1 is deeper. These deeper troughs that exist downstream of z = 0.405 mas cannot be achieved with these axisymmetric models. The same situation is true with axisymmetric models of extreme limb brightening in planetary nebulae. It has been proposed that fortuitous asymmetric structure is required for extreme limb brightening in planetary nebulae \citep{mas90}. This might be the case for M87 as well. The extra degrees of freedom implicit in this scenario are not desirable to introduce here as the effort is focused on understanding the basic geometry that is responsible for a persistent prominent phenomenon of deep troughs and wide ridges. Speculation on the asymmetries required in the jet wall will be left for future research.

\par The investigation began with the study of simple cones and cylinders. The wall widths of the cones and cylinders were varied and the opening angle of the cone was varied as well. The axisymmetric emissivity function was also allowed to vary. No model could even produce the shallowest troughs in the 43 GHz cross-cuts from \citet{pun22}, i.e.that which is shown in the top left frame of Figure 2. The parameter $W/R$ in of itself cannot produce these deep troughs. The general trends are summarized in the top left panel of Figure 2. Choosing a value smaller than $W/R = 0.1$ does not deepen the trough very much due to the relatively low resolution of the restoring beam and it hinders the ability to create a wide ridge as indicated in the top right hand panel of Figure 2. The models require another degree of freedom in order to succeed. The model that produces the deepest troughs in the top left panel of Figure 2 has, $W/R$, that decreases linearly towards a conical apex as in the bottom panel of Figure 2 (``tapered cone", hereafter).

\par The 3-D spatial relationships are rather complex and the bottom panel of Figure 2 is provided as a visual aid. The dark, olive green region is where the LOS, along a cross-cut, intersects the wall of the tubular jet. The LOS in the middle of the cross-cut at x=0 slices through two regions that are very far apart. Since, M\,87 has a deep trough and the emissivity is highest close to the nucleus, the tapering of the cone wall helps achieve a deep trough by decreasing the area of the intersection of the blue LOS plane with the jet wall on the far right, where the emissivity is high. Table 1 indicates a nearly cylindrical jet for $z = 0.4-0.65$ mas, so our fiducial geometry is a tapered cone at the base and a cylinder at larger z. A $28^{\circ}$ cone intrinsic opening angle is chosen \citep{wal18}. In summary, the additional required parameter that creates the deep troughs is the amount of wall thickness tapering in the conical region. The variable, $W/R$, at the junction of the cylinder and the cone cannot create the required trough depth. However, the conical wall taper can produce troughs representative of those in the 86 GHz cross-cuts for a wide range, $W/R \leq 0.35$. If $W/R$ tapers by a factor of $\sim 4$, the troughs of the 86 GHz VLBI cross-cuts can be achieved as in the top right hand frame of Figure 2. The method is limited by the relatively low resolution of VLBI and cannot produce the deeper troughs seen at 43 GHz (like the one in the top right hand panel of Figure 1) regardless of the rate of the linear tapering of the wall thickness.

\begin{table}
\begin{center}
 \caption{Jet Properties of the Image in Figure 1 \citep{pun22}}
\begin{tabular}{cccc}
 \hline
 $z$  & $z'\equiv$ the De-projected & Distance & Flux Density \\
 & Axial & Between & of 0.16 mas Wide \\
 & Displacement & Intensity Peaks & Cross-Section \\
  (mas)& (mas) & (mas)   & (mJy) \\
 \hline
0.405 & 1.31 &$0.514\pm0.040$ & $66.4 \pm 6.7$\\
0.45 & 1.46 &$0.555\pm0.040$ & $52.2 \pm 5.3$\\
 0.5 & 1.62 & $0.542\pm0.040$ & $44.6 \pm 4.6$ \\
 0.55 & 1.78 &$0.548\pm0.040$ & $33.5 \pm 3.5$\\
0.6 & 1.94 &$0.518\pm0.040$ & $32.2 \pm 3.4$ \\
0.65 & 2.11 &$0.523\pm0.040$ & $33.2 \pm 3.4$
\end{tabular}
\end{center}
\end{table}
\par The top left panel of Figure 2 indicates a degeneracy in parameter space: by just making the walls very thin in the tapered region of the cone, one can get troughs of depths similar to the z = 0.405 mas cross-cut, regardless of W/R. However, the wider ridges adjacent to these troughs are a definite challenge to comprehend. The top right panel of Figure 2, studies the effect of the wall thickness of the model on the observed ridge width. The models under consideration are axisymmetric, so the study utilizes the most symmetric cross section of the 86 GHz image in which the ridges are resolved (z = 0.55 mas) for analysis in the upper right hand panel of Figure 2. Since the restoring beam is relatively large, the convolved ridge thickness increases slowly with increasing $W$ and the trough gets shallower (see the top right of Figure 2). The best description of the cross-cuts has $W/R=0.35$ at the junction of the cylinder and cone and tapers to $W/R=0.08$ when the LOS exits the cone wall at x=0 (abbreviated as $0.08 < W/R < 0.35$, hereafter). But, the ridges are not as wide as observed. If one chooses $W/R=0.40$, the trough depth requirement will be breached at the expense of a little more observed ridge width. This is the essence of the conflict of wide ridges adjacent to deep troughs, the fundamental conundrum presented by the observations. These simplest models are close to observation, but not sufficient.

To gain additional intensity ridge width, without raising the trough substantially, a sheath (i.e., a dissipative shear layer between the external medium and the thick-walled tubular jet) was wrapped around the tubular jet. It has the same shape as the tubular jet, the inner profile of the sheath is the same as the outer profile of the tubular jet. It is therefore a wider version of the geometry in Figure 2, with narrow walls and a taper abbreviated as $0.03 < W/R < 0.1$. This is successful as a perturbation because the geometric source for the weak sheath is wider by construction, so it is more resolved by the convolving beam. This accentuates the difference between long and short path lengths through the jet wall, i.e. it deepens the trough. This also explains the deeper troughs as the physical jet expands in the bottom right of Figure 1 and Figure 3 which shows the resulting fits. Since the models are axisymmetric by construction, they cannot reproduce a cross-cut with ridges of different peak intensity. Hence, we introduce the cross-cut at z = 0.35 mas in the left hand panel of Figure 3, even though the jet is too narrow to be fully resolved by the VLBI experiment. The most symmetric cross-section that is transversely resolved is at z = 0.55 mas and the corresponding fit is shown on the right hand side of Figure 3. Notice that the sheath is merely a small perturbation of the thick walled jet in the top panels. However, the fit is not unique. If one chooses a smaller value of the wall thickness in the tubular jet, say $W/R=0.1$ for the sake of argument, then the contribution of the sheath must increase considerably in order to fill-in the wide wings of the ridges (see the bottom panel of Figure 3). It is no longer just a perturbation. In order to achieve the required limb width there is a gap of $\sim 0.105$ mas between the intensity peak of the sheath and intensity peak of the wall. Since the sheath is of comparable surface brightness to the jet wall, the model of the emission is the aggregation of the two tapered cones plus cylinders and the gap. One can compute the total thickness of the emissivity region to get $W_{\rm{total}}/R = 0.37$. A wide emission region surrounding a very low luminosity core seems inescapable regardless of the partitioning into different labels. The gap is not impossible, but an adjacent sheath is more compatible with an enveloping shear layer. The choice here is to make the jet wall as close to the complete solution as possible in order to minimize the contribution of the outer sheath, i.e., make it a perturbation representative of an adjacent shear layer. These results do not prove an exact model this was never the intent of the simplified axisymmetric models), but they indicate that a very wide structure of a tubular wall and surrounding sheath are needed to create the wide ridges. An approximate estimate of the ratio of the tubular jet volume to the total jet volume is $\geq 1-(1-W/R)^{2}=0.58$ at z = 0.35 mas and z = 0.55 mas.
\begin{figure}
\begin{center}
\includegraphics[width= 0.47\textwidth]{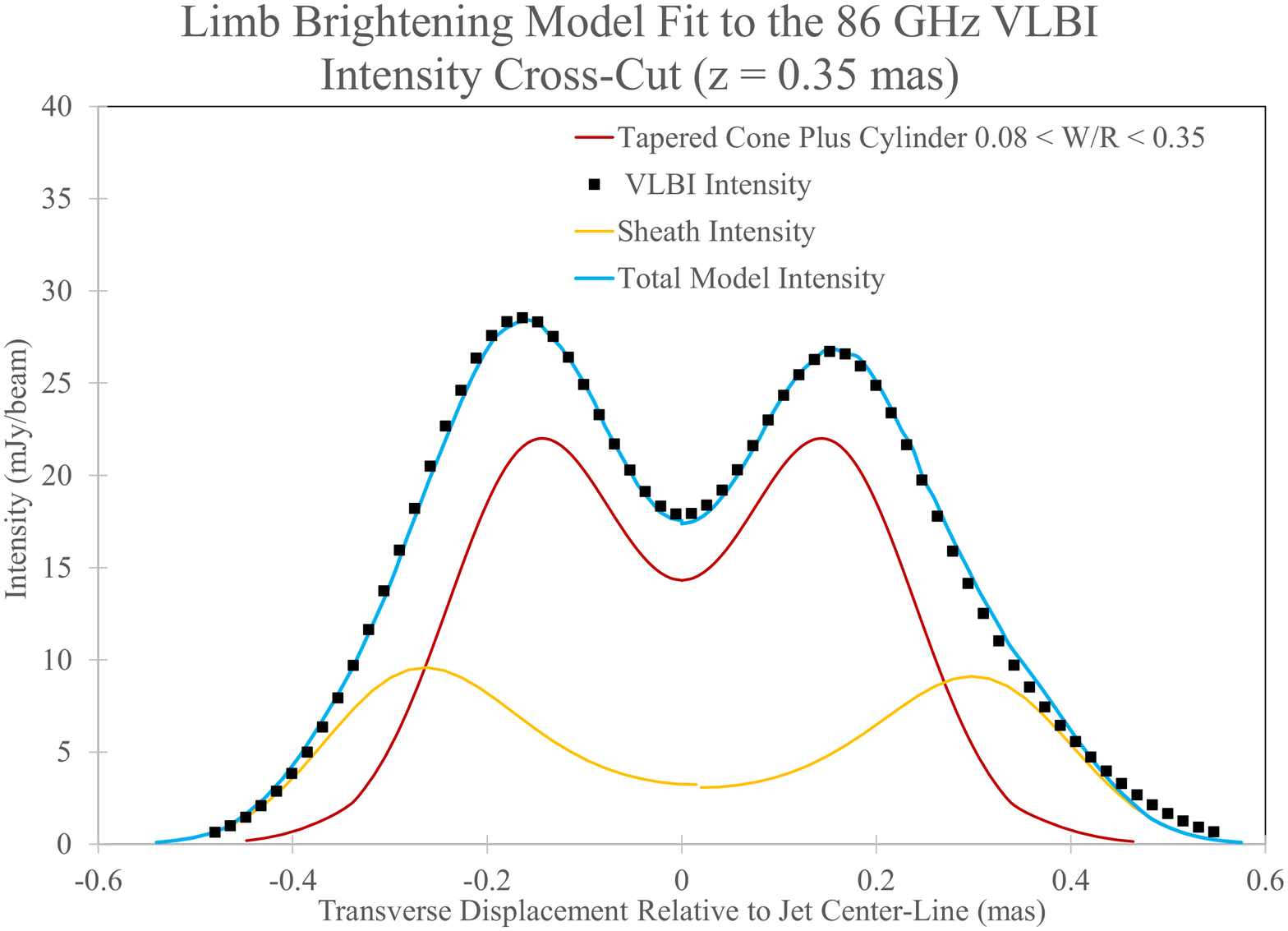}
\includegraphics[width= 0.47\textwidth]{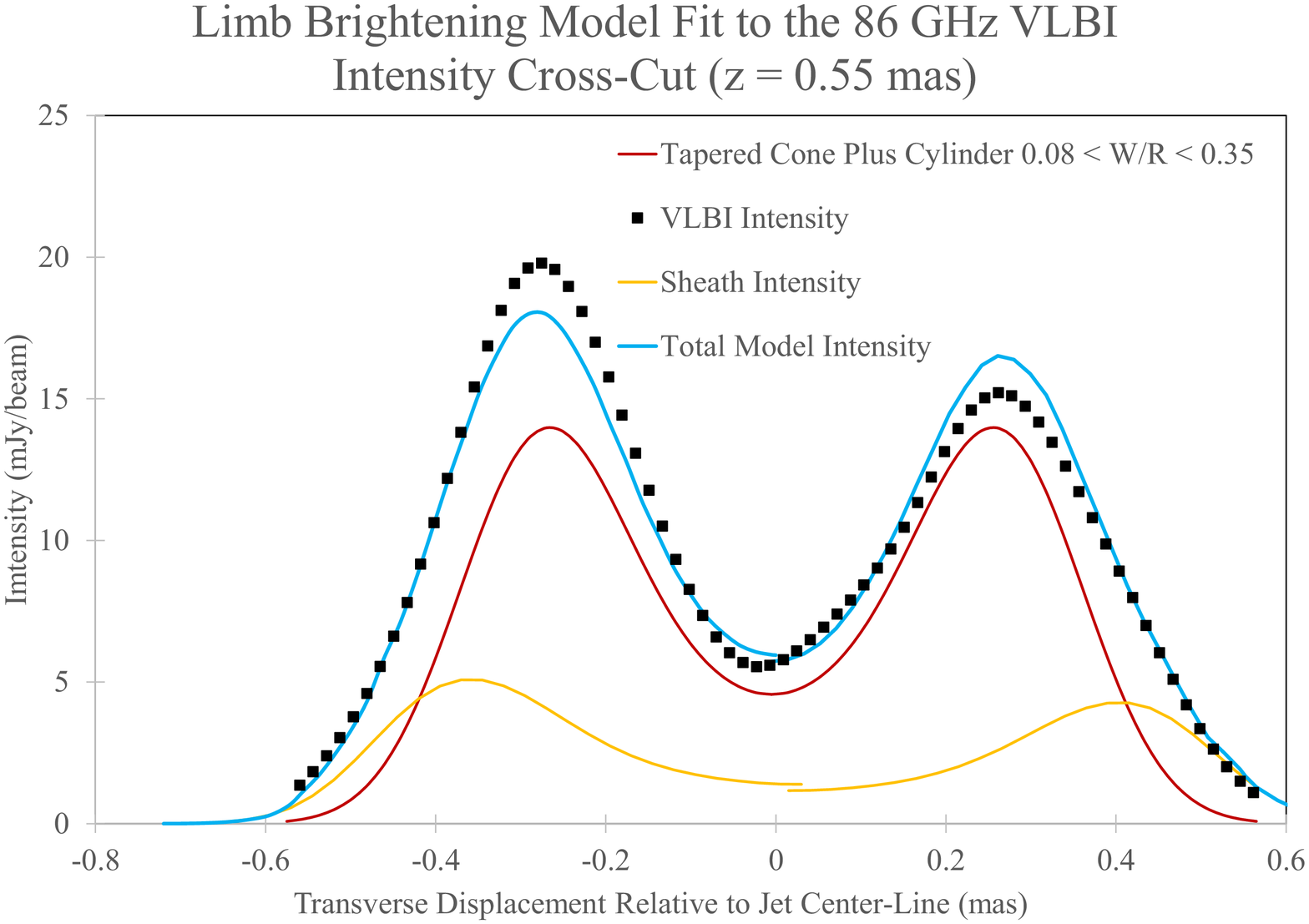}
\includegraphics[width= 0.47\textwidth]{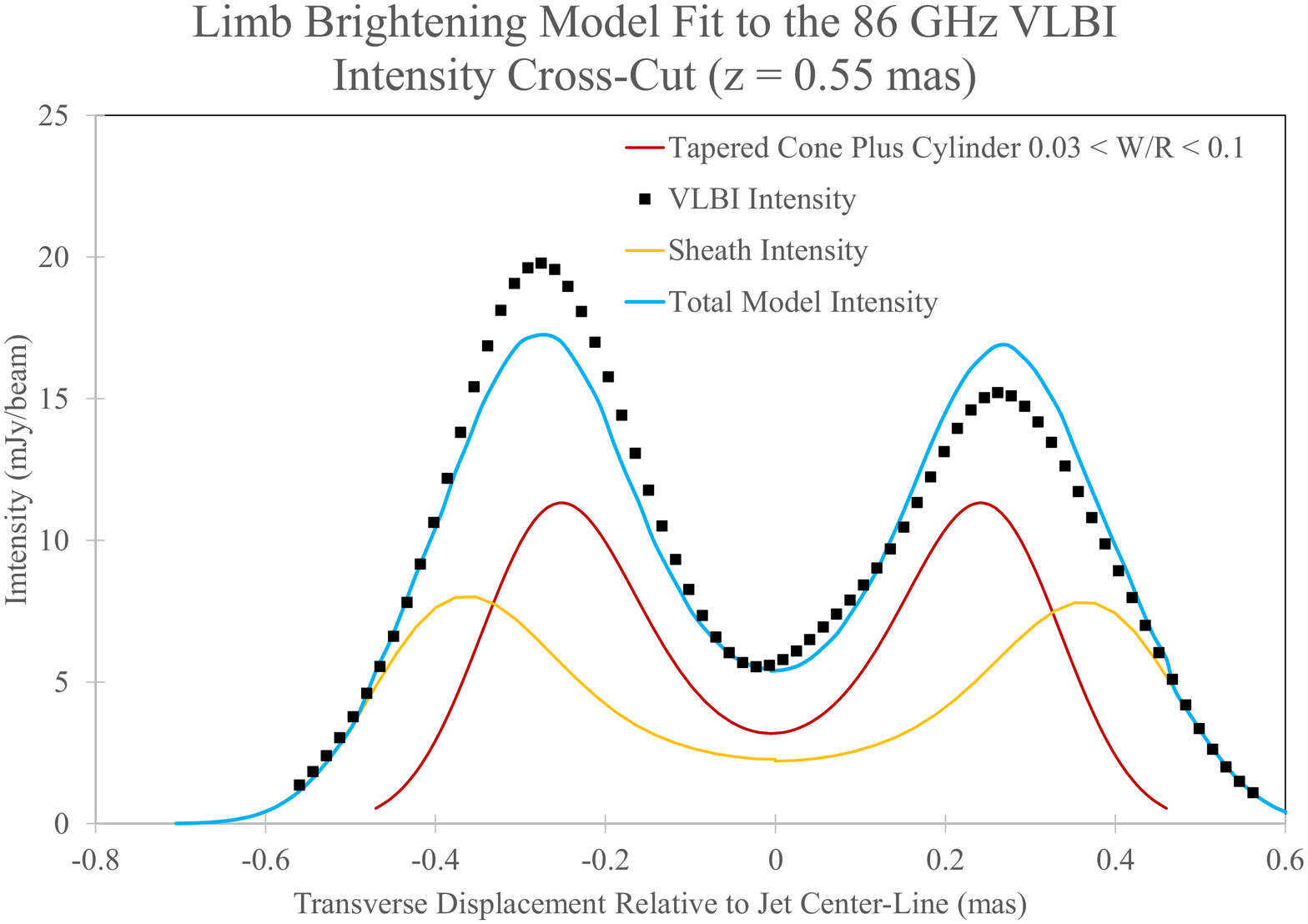}
\caption{\footnotesize{The top panels of the figure show the fit to two intensity cross-cuts. There are two components: the main body of the jet, a tapered cone plus cylinder ($0.08<W/R<0.35$) and a sheath, an outer tapered cone plus cylinder ($0.03<W/R<0.1$). The cross-cut on the top left is the most symmetric cross-cut from the 86 GHz VLBI image and is the most compatible with an assumed axisymmetric model. The lack of symmetry between the north and south ridges in the cross-cut on the top right renders a single axisymmetric model imperfect. The bottom panel is an alternate method of implementing the model. In this model, the jet wall thickness is narrow ($0.03<W/R<0.1$). This fit has comparable surface brightness contributions from the putative jet and the sheath.}}
\end{center}
\end{figure}
\par It is worth noting that one can also consider the cross-cuts of the super-resolved version of the 86 GHz image in the bottom left frame of Figure 1 that was previously published \citep{had16,pun18}. The image was restored with a circular beam of 0.11 mas instead of the 0.116 mas x 0.307 mas, PA=$-9^{\circ}$ restoring beam of Figure 1. The axis that is parallel to the major axis of the synthesized beam is heavily super-resolved (by $\approx 30\%$). However, there are no baselines that are long enough to resolve objects as small as 0.11 mas along the major axis of the synthesized beam. Thus, results based on an individual feature of the super-resolved image are not as reliable as those based on the source reconstruction in Figure 1. The results are only suggestive taken as a whole and certainly not very reliable taken individually. The relevant cross-cuts of the super-resolved image from z= 0.3, 0.4 and 0.5 mas were previously published \citep{pun18}. The main advantage of a high resolution image is the possibility of achieving a stronger connection between the FWHM of the de-convolved Gaussian fits of the peaks, $\rm{FWHM}_{\rm{deconvolved}}$, and the intrinsic width of the jet walls. Thus motivated, the six intensity peaks from the three cross-cuts were fit with Gaussian functions then de-convolved from the circular restoring beam. $R$ is defined as one-half the distance between the outer de-convolved half maxima of the two ridges. It is found that the mean value of $\rm{FWHM}_{\rm{deconvolved}}/R=0.41$ and the median value is $\rm{FWHM}_{\rm{deconvolved}}/R=0.35$. This is consistent with the conclusions presented above.
\section{Simulated Jets Contrasted with the Sub-mas Jet in M\,87}
It was previously noted that synthetic images of numerically simulated jets (typical of those in the EHT Collaboration library) were much narrower (about a factor of $0.4-0.7$) compared to 86 GHz VLBI observations of jets for $z\leq 0.25$ mas \citep{pun19,cru21}. However, the apparent non-existent limb brightening for $z<0.6$ mas could not be quantified in an easy to grasp format since the raw data of the simulations were not made public \citep{pun19,cha19,mos16}. Fortunately, publicly shared synthetic image (Flexible Image Transport System, FITS) files generated from general relativistic simulations were recently created with the same restoring beam as the 86 GHz VLBI observation in Figure 1 \citep{cru21}.

The lower right panel of Figure 1 shows intensity cross-cuts extracted from the synthetic image FITS file compared to cross-cuts the same distance down the jet axis for the 86 GHz VLBI observation. The simulated jets have a peak in the middle and the VLBI jet has a deep trough in the middle. The simulated jet also fades too rapidly with z, both properties appear in other published numerically simulated synthetic images \citep{mos16,cha19,pun19,fro21}. The EHT Collaboration have assessed their large library of simulations and have determined that only their simulations of magnetically arrested accretion (MAD) can explain the coarse properties of the annulus and produce a jet with enough power \citep{eht19,eht21}. It has been shown that high spin black holes, $0.5<a/M<1$ (where $Ma$ is the angular momentum of the black hole in geometrized units: black holes are only defined for $-1<a/M<1$) produce the widest MAD jets and are therefore the least nonconforming with the observed jet width, noted above \citep{cru21,fro21}. The MAD, a/M=0.94 simulations have a brighter jet than the MAD, a/M=0.5 simulations \citep{cru21}. Considering the rapid fading of the simulated jet (in the lower right panel of Figure 1) compared to the VLBI jet, the a/M=0.94 simulations are closer to reality. Thus, the simulation presented in Figures 1 and 4 was chosen for maximum conformance to observation.
\par The highest emissivity in the simulated jets occurs in a narrow, surrounding ``sheath" adjacent to an invisible (by construction) jet. Edge brightening (only the edge of the jet is visible) is maximized by construction, but limb brightening on the sky plane is never pronounced in synthetic images \citep{mos16,cha19,cru21,fro21,yan22}. To understand this apparent contradiction, consider the left hand side of Figure 4 which shows the source function before convolution with the restoring beam. We superimposed ellipses representing the restoring beam FWHM at the locations of the peak surface brightness (in Figure 1) for three cross-cuts. Notice that these peaks are not due to LOS geometry and limb brightening. They simply represent local emissivity maxima on a background of low emissivity (almost voids in brightness). These bright areas arise from a low filling factor set of high dissipation regions (perhaps shocks). If the tubular jet had thicker walls and a higher median emissivity, the jet walls would provide a much larger Gaussian weighted emissivity column within the restoring beam. As such, the bright areas would not dominate and limb brightening would be conspicuous on the sky plane. As expected, a parameter study of emissivity functions and locations of boundaries between the sheath and the invisible leptonic jet never changes this wafer thin appearance of the sheath \citep{fro21}. Thus, near the jet base, the limb brightening on the sky plane will never be strong in this family of simulations. The absence of limb brightening in the sly plane in the synthetic images is more evidence supporting the conclusion of the analysis in the previous section, the jet walls in M\,87 are very thick.
\begin{figure}
\begin{center}
\includegraphics[width= 0.47\textwidth]{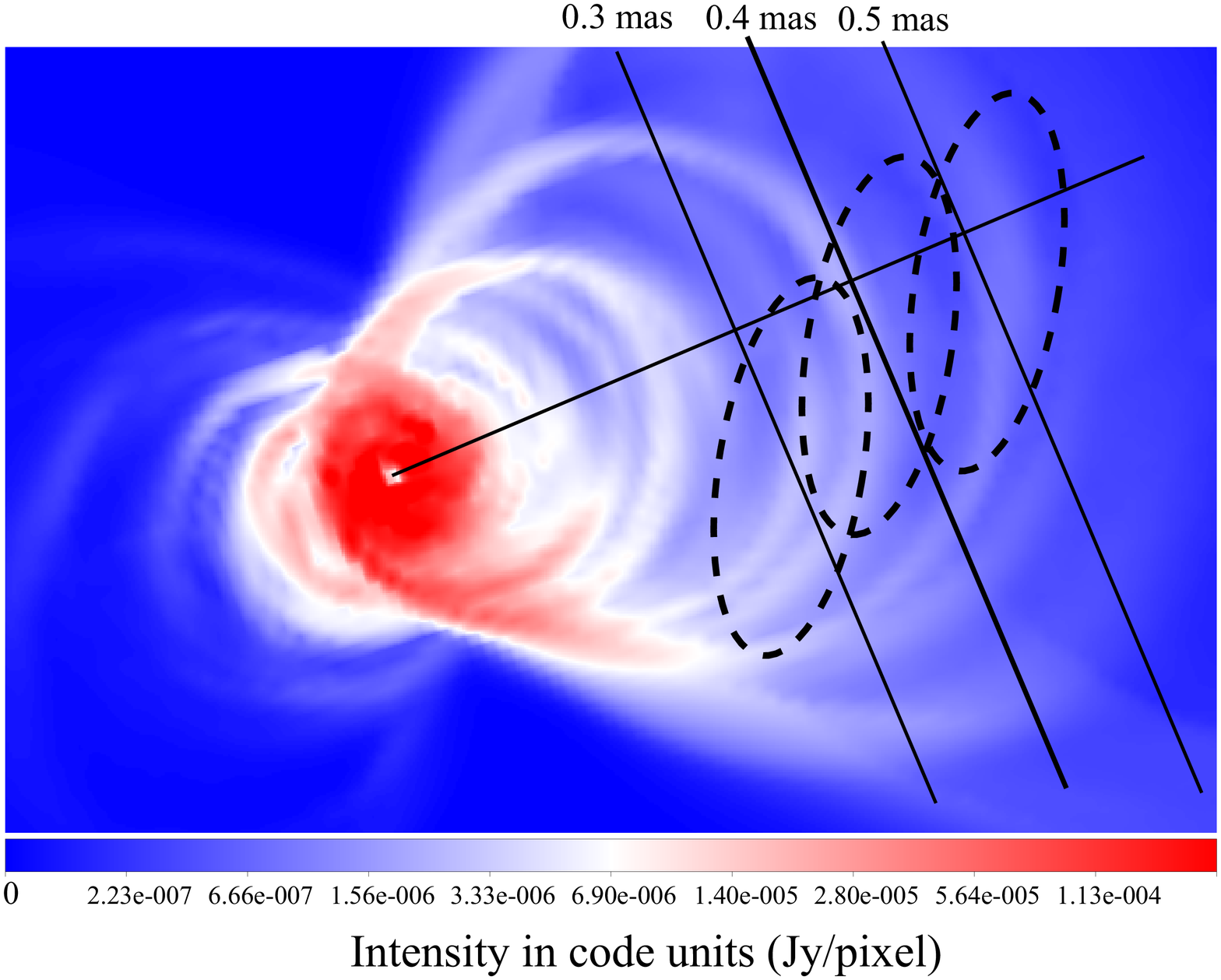}
\includegraphics[width= 0.47\textwidth]{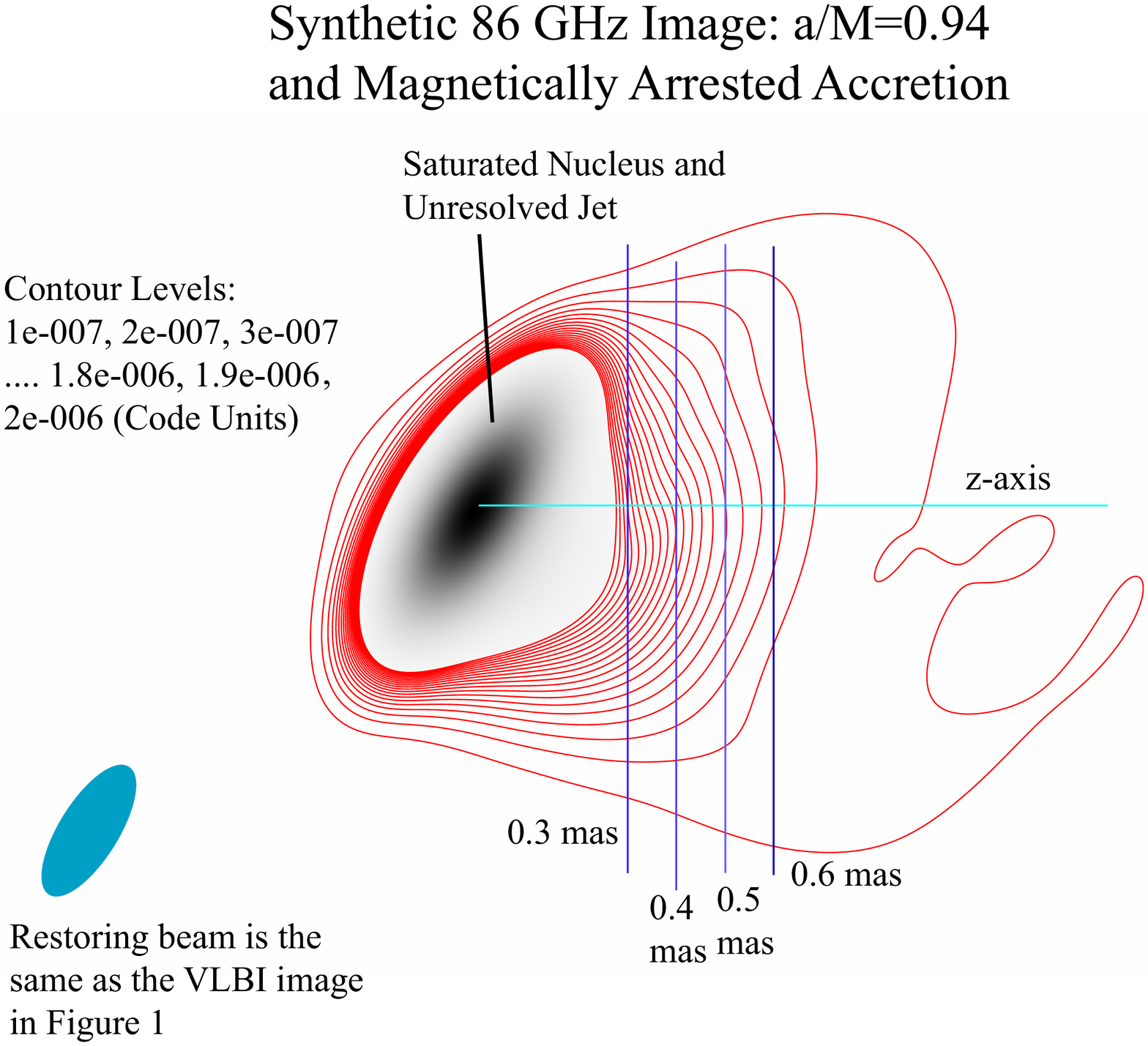}
\caption{\footnotesize{The source function (left) for the synthetic image on the right provides some clues to the lack of limb brightening on the sky plane. The dashed black ellipses are the FWHM of the restoring beam at the peak intensity of the cross-cuts in Figure 1. The locations of the cross-cuts are indicated in the image on the right \citep{cru21}. The bright peaks are just local bright spots on an otherwise mostly faint, hollow, conical jet. There is no significant geometrical limb brightening on the cross-cuts.}}
\end{center}
\end{figure}

\section{Conclusion}
The analysis of limb brightening determined that the source of the observed jet base emission in M\,87 is a tubular jet with more than half of the total jet volume. If the jet were leptonic and magnetically dominated, being beyond the light cylinder \citep{pun22}, the particles would flow at highly relativistic velocities \citep{mck06}, contrary to the mild Doppler boosting inferred from the top left image in Figure 1. This is consistent with the unique magnetohydrodynamic solution consistent with that image. It was determined that the only outflows that satisfy the mildly relativistic bulk flow velocity profile, the flux density distribution along the jet as well as the conservation laws of mass, energy, and angular momentum are tubular jets of protonic plasma driven by a large scale magnetic field with a significant Poynting flux \citep{pun22}. The energy flux at this epoch is $\gtrsim 10^{42}$ erg/sec and the wind transports $\gtrsim 10^{-4}$ solar masses per year \citep{pun22}.

\par The results of this study can be used as guidelines to modify the numerical simulations when considered from a comprehensive perspective. There are three issues that are inter-related and need to be considered simultaneously,
\begin{enumerate}
\item At a de-projected axial distance of $\sim 100$M  – 200M from the nucleus, the synthetic images of simulated high black hole spin, MAD, jets are approximately $\sim 0.4-0.7$ as wide as the jet in M87 \citep{pun19,cha19,cru21}. This implies that the luminous jet streamlines in simulations need to diverge more rapidly as the jet propagates away from the launch site in order for jet width to be consistent with the observed transverse structure.\footnote{A new simulation claims to have resolved this issue \citep{yan22}. However, since the jet is very center brightened, they compare the FWHM of the of the central ridge in the simulations to the ridge to ridge separation in the ``double rail" images. This is not an equitable method of comparison. If the methods of \citet{pun19} are applied to these same synthetic images, no width difference is seen at $z<0.45$ mas from the synthetic images of \citet{cha19}. So this issue seems not to be resolved.}
\item 300 M - 550 M downstream of the nucleus, the jet is extremely limb brightened and the simulated jets are center brightened.
\item The extreme limb brightening and wide ridges can only be made by a very thick-walled tubular jet with over half the total jet volume.
\end{enumerate}
A conversion of 3.8 $\mu$as = 1M and $M\approx 9.6\times 10^{14}$ cm was assumed in the above \citep{eht19}.

\par One possible resolution is to extend the emissivity into the jet to fix points 2) and 3). This is a scientific and computational issue because the mass flux is set by an invoked mass floor and numerical diffusion in the jet \citep{cha19}. Also, this would make the jet appear narrower, therefore more nonconforming to point 1). Furthermore, it would likely create a bulk flow velocity that exceeds the mildly relativistic estimates based on Figure 1 \citep{mck06,pun21}. Alternatively, one might want to extend the emission region outward into the protonic region. The only collimated protonic outflows in the EHT Collaboration library of simulations are the ``funnel wall jets" created as the accreting gas rams into a centrifugal barrier and is accelerated by pressure in the accretion disk's corona \citep{por19,haw06}. This is a very thin outflow (perhaps much of what we see in Figure 4). There is also the possibility of a magneto-centrifugally driven wind from the disk. But there are few organized large scale polodoidal fields in the disk in these simulations. It is mainly chaotic (turbulent) with a net poloidal excess when averaged over time and azimuth, these are not the properties that lead to a collimated jet (probably the reason for its absence in synthetic images) \citep{tch15,bla82,qia18}. Another possibility is that there is a huge amount of entrainment in this first 0.25 lt-yrs of jet propagation ($\sim 10^{-4}$ solar masses per year) that is not captured by the simulations. Not only would there need to be large amounts of entrainment, but it would need to migrate 1/3 of the distance to the jet center. For geometrical reasons (larger lever arm), these are the streamlines with most of the created jet Poynting flux power in the simulations \citep{kro05}, so the interior smaller volume would carry the minority of the jet power. This would contradict the notion that most of the jet power is hidden in an invisible core or spine \citep{ghi05}. This scenario does not remedy the non-conformance in point 1).

This study benefitted from the comments of a helpful referee. I am indebted to Sina Chen for the intensity cross-cuts. Christian Fromm generously provided the image FITs files used in this paper from Cruz-Osorio et al.(2021). Alan Marscher and Sam Gralla provided comments on earlier versions of this manuscript that greatly improved the science.

\end{document}